\documentclass[aps,12pt,reprint]{revtex4-1}
\usepackage{graphicx}
\usepackage{epsfig}
\usepackage{epsf}
\usepackage{subfigure} 
\usepackage{epstopdf} 
\usepackage{color}
\usepackage{amsmath} 
\usepackage{dcolumn}
\usepackage{bm}
\usepackage{amssymb} 
\usepackage{mathtools} 
\usepackage{subeqnarray}
\usepackage{mathrsfs} 
\usepackage[colorlinks=true, pdfstartview=FitV,
  linkcolor=red, citecolor=blue, urlcolor=blue]{hyperref}
\usepackage{natbib} 
\usepackage{url} 
\usepackage[utf8]{inputenc}
\usepackage[english]{babel}
\usepackage{float}

\newcommand{\be}{\begin{equation}} \newcommand{\ee}{\end{equation}}
\newcommand{\ben}{\begin{eqnarray}} \newcommand{\een}{\end{eqnarray}}

 \newcommand{\bsen}{\begin{subeqnarray}}
\newcommand{\esen}{\end{subeqnarray}}

\begin{document}
\title{Tomlinson model improved with no ad-hoc dissipation}

\author{María Luján Iglesias}\email{lujan.iglesias@ufrgs.br} \affiliation{Instituto de
  Física, Universidade Federal do Rio Grande do Sul, Porto Alegre, RS,
  Brazil} \author{Sebastián Gonçalves}\email{sgonc@if.ufrgs.br} \affiliation{Instituto
  de Física, Universidade Federal do Rio Grande do Sul, Porto Alegre,
  RS, Brazil}

\begin{abstract}
The origin of friction force is a very old problem in physics, which
goes back to Leonardo da Vinci or even older times. Extremely
important from a practical point of view, but with no satisfactory
explanation yet. Many models have been used to study dry sliding
friction. The model introduced in the present work consists in one
atom that slide over a surface represented by a periodic arrangement
of atoms, each confined by an independent harmonic potential. The
novelty of our contribution resides in that we do not include an {\em
  ad hoc} dissipation term as all previous works have done.  Despite
the apparent simplicity of the model it can not be solved
analytically, so the study is performed solving the Newton's equations
numerically.  The results obtained so far with the present model are
in accordance with the Tomlinson model, often used to represent the
atomic force microscope. The atomic-scale analysis of the interaction
between sliding surfaces is necessary to understand the
non-conservative lateral forces and the mechanism of energy
dissipation which can be thought as effective emerging friction.
\end{abstract}

\maketitle

\section{Introduction}
Friction is one of the most important problems and a key phenomenon in
physics and engineering, whose fundamental origin has been studied for
centuries and still remains controversial
~\cite{Krim2002,Persson2000,Makkonen2012}. From Leonardo da Vinci,
Coulomb and Rynolds, sliding friction has been a topic of great
interest studies intensively by many of the brightest scientists. In
the last years, theoretical models for atomic friction, mostly based
on the early work of Tomlinson ~\cite{Tomlinson1929}, and Frenkel-Kontorova~\cite{Kontorova,Frenkel,braun2004frenkel} models, were proposed and characterized by being highly simplified and yet retaining enough
complexity to exibit interesting features
~\cite{Buldum1997,Goncalves2004,Goncalves2005,Tiwari2008,Fusco2005,Neide2010}.
Such models have allowed to explain essential features of atomic-scale
friction such as the occurrence of the "stick-slip" phenomenon
observed in the movement of the tip over a surface material in the
friction force microscope (FFM) ~\cite{Tomanek1991,Gyalog1996}.
Friction is the result of the transformation of sliding motion into
heat or different forms of energy, i.e., is the dissipated energy per
unit length ~\cite{Helman1994,Wang2015}, where the frictional work it
is assumed to be dissipated through plastic deformation and material
damage ~\cite{Bowden1951}.  In the Tomlinson model, the energy
dissipation is attributed to the instability induced by the stick-slip
motion and consequent atomic vibration.  What is expected is to
increase the control over the mechanisms of friction and thus reduce
the loss of energy.  In this contribution we aim to understand the
fundamental mechanisms on these energy exchange between the tip and
the surface without include any {\em ad hoc} dissipation term as
previous works have done. We want to show that without the {\em ad
  hoc} term, the energy lost by the tip is absorbed by the substrate.
Despite the apparent simplicity of the model it can not be solved
analytically, so the study is performed solving the Newton's equations
numerically.  demonstrate that the energy dissipation is almost all
absorbed by the stiffness of the substrate

\section{Model}

\begin{figure} 
\centering \includegraphics[width = 0.90\columnwidth]{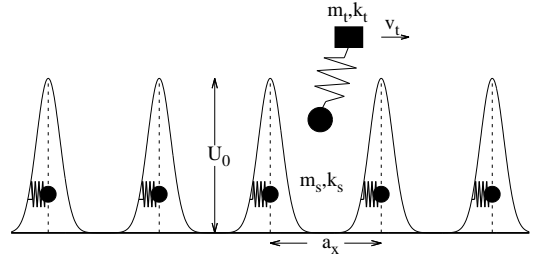}
\caption{Diagram of the model proposed in this
  work.} \label{fig:model}
\end{figure}

The proposed model in the present study aims to explore the effect of various parameters involved on the emergence of frictional force and how the energy generated during the process is dispersed when there is no \textit{ad hoc} dissipation. 
In this sense, our model in Fig.~\ref{fig:model} was never presented in previous works.
The tip is represented by a particle of mass $M$, connected by a spring of constant $K$ to a driven support that moves at constant velocity $v_{c}$.
The substrate is represented by a series of particle-spring systems of mass $m$ and constant $k$, independent between them , but interacting with the tip via a short range Gaussian type potential. To avoid edge
effects we modeled the chain as being infinite.

 The system is described by the Hamiltonian \be H = \frac{P^{2}}{2M}
 +\sum_{i=1}^{N} \frac{p^{2}}{2m} + \sum_{i=1}^{N} U(X,x_{i}) + U(X) +
 \sum_{i=1}^{N} U(x_{i}) \ee

where $X,P$ and $x,p$ are the tip and substrate coordinates
respectively; $U(X,x_{i}) = U_{0}e^{-\frac{(X-x_{i})^2}{\sigma^2}}$ is
the interaction between the tip and each of the particles of the
substrate.  We choose the value of $\sigma$ in such a way that the 
Gaussian potential of each particle does not overlap with that of its neighbors.
So the tip can interact with each of them
separately.\\
In this case, $\sigma = a_{x}/n$, where $a_{x}$ is the separation
between the particles of the substrate and $n$ is a constant.  $U(X) =
\frac{K}{2}(X - x_{c})^2 $ is the elastic interaction between the tip
and the cantilever.  The support position $x_{c}$ is $x_{c} =
v_{c}t$. And $U(x_{i}) = \frac{k}{2}(x_{i} - x^{0}_{i})^2$ is the
elastic potential of each particle (independent between them).  

This results in the following equation of motion:

\ben
\begin{aligned}
 \begin{multlined}
\label{Eq:newton_p}
 M\ddot{X} = -K(X - v_{c}t) \\ +
 \frac{2U_{0}}{\sigma^2}\sum_{i=1}^{N}(X - x_{i})e
 ^{-\frac{(X-x_{i})^2}{\sigma^2}} \\
\end{multlined}
\end{aligned}
\een \ben
\begin{aligned}
\begin{multlined}
\label{Eq:newton_subs}
m \sum_{i=1}^{N} \ddot{x_{i}} = - k\sum_{i=1}^{N}(x_{i}-x_{0_{i}})\\ -
\frac{2U_{0}}{\sigma^2}\sum_{i=1}^{N}(X - x_{i})e
^{-\frac{(X-x_{i})^2}{\sigma^2}}
\end{multlined}
\end{aligned}
\een

Introducing the adimensional units $ Q = \frac{X}{\sigma} , q =
\frac{x}{\sigma}, \tau = \sqrt{\frac{K}{M}} t , \tilde{U_{0}} =
\frac{2U_{0}}{k_{t}\sigma^2} , \tilde{v_{c}} = \frac{v_{c}}{\sigma}
\sqrt{\frac{M}{K}}$ , the equations ~\ref{Eq:newton_p} and
~\ref{Eq:newton_subs} can be reduced to:

\ben \ddot{Q} = -q + \tilde{v_{c}}\tau +
\tilde{U_{0}}\sum_{i=1}^{N}(Q-q_{i})e^{-(q_{t}-q_{s_{i}})^2} \een \ben
\begin{aligned}
\begin{multlined}
\sum_{i=1}^{N} \ddot{q}_{i} = \frac{\epsilon_{1}}{\epsilon_{2}}
\sum_{i=1}^{N} (q_{i} - q_{0_{i}})\\ -\epsilon_{1} \tilde{U_{0}}
\sum_{i=1}^{N}(Q-q_{i})e^{-(Q-q_{i})^2}
\end{multlined}
\end{aligned}
\een where $\epsilon_{1} = M/m$ is the ratio of the mass of the tip
and the substrate particles (assuming all have the same masses) and
$\epsilon_{2} = K/k$ is the ratio of the stiffness of the tip and the
substrate. 

\begin{figure}[!h]
\centering \includegraphics[width =
  1.0\columnwidth]{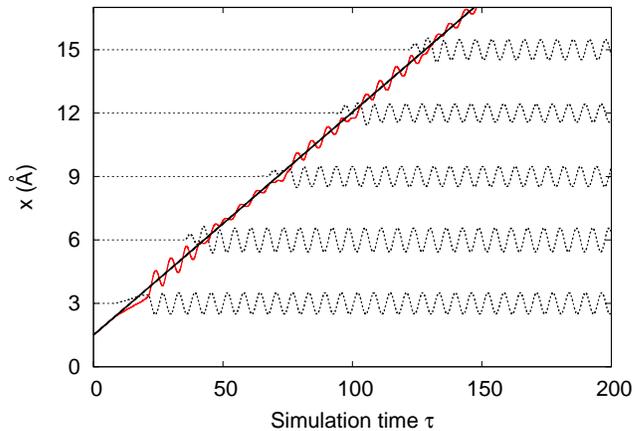}
\caption{Position of the tip and the cantilever guide in straight line. The dashed lines represents the first five perturbed particles. }.\label{Fig:pos}
\end{figure}

\section{Methods}

The problem was solved numerically by using classical molecular-dynamics methods ~\cite{Allen1987,Frenkel2002}
using a set of realistic parameters~\cite{Zworner1998,Fusco2005,Wang2015}, $M = m = 10^{-10}$ $kg$, $K =
k = 10$ $N/m$, $a_{x} = 3$ $\AA$, $U_{0} = 0.085$ $eV$, $v_{c} = 1$
$\mu m/seg$ which are typical of an AFM experiments ~\cite{Bennewitz1999,Holscher1997}.\\
The most evident consequence of not having dissipation is that the particles that are perturbed by the tip do not return to their initial state, remaining oscillating infinitely, as it is presented on Fig.~\ref{Fig:pos} for the first five particles in dashed lines.\\
For the frictional force calculation we compare three different methods.

It is known that $F$ is defined
as the mean value of the lateral force
~\cite{Holscher1997,Zworner1998,Fusco2005,Helman1994}. Since the
problem is in one dimension, from here, we will refer to the lateral
force as friction force $F$.
\ben 
F = K(X-x) = K(X-v_{c}t) 
\een 

The
second way to obtain $F$ is through the energy
accumulated by the substrate.  The power is the instantaneous product
of force times velocity and the time derivative of energy.  
\ben
F\ v_{c} = \frac{dE}{dt} 
\een

\begin{figure}[!h]
\centering \includegraphics[width =
  1.0\columnwidth]{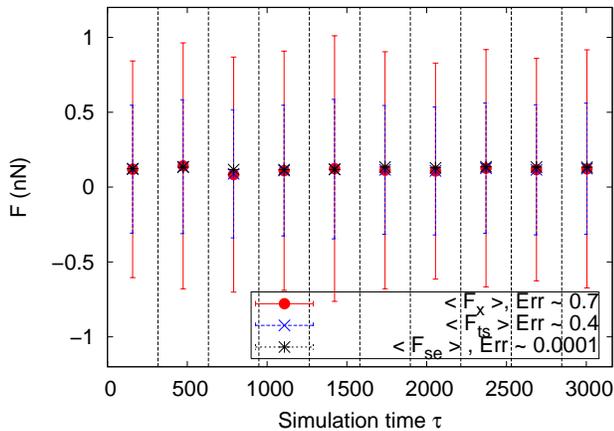}
\caption{Comparison of the friction force values for the three
  methods of calculation, and their respective error bars}.\label{Fig:av_forces}
\end{figure}

From this relation, it is possible to obtain the frictional force as
the slope of the total substrate energy by linear regression. 
 Another method to calculate the friction force in these mechanical problem is
calculating the average of the force between the tip and the substrate
generated by the gaussian interaction force.  These three methods are
compared in Fig.~\ref{Fig:av_forces} with their respective error bars.

The total simulation time $\tau$ was divided into ten equal regions.  For each of these intervals, the mean of the force was calculated in order to compare their values.
$<F_{x}>$ represents the value obtained from
eq. ~\ref{Eq:newton_subs}, $<F_{ts}>$ from the gaussian potential
between the tip and the substrate; and $<F_{se}>$ is the slope of the
accumulated energy of the substrate.

\begin{figure}[!h]
  \centering \includegraphics[width =
    1.0\columnwidth]{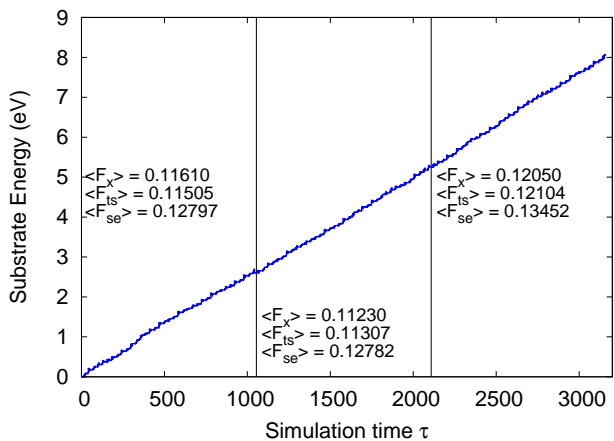}
  \caption{Comparison of the lateral force values for the three
    methods in diferents times}.\label{Fig:fx_dt10}
\end{figure}

In Fig.~\ref{Fig:fx_dt10} it is shown the accumulation of energy by
the substrate and the values of the frictional force obtained by the
three methods. In this case, $\tau$ was divided into three regions.

The comparison of the three methods shows that the
smallest error in the measurements corresponds to the force obtained
through the energy accumulated by the substrate. 
For this reason, we have chosen it to perform the rest of the simulations where all the parameters of the model will be analyzed. 
\subsection{Time step comparation}

Molecular dynamics simulation is a technique by which, step by step, the equations of motion that describe the Newton's classical mechanics are solved ~\cite{Allen1987,Rapaport1995,Frenkel2002}.
In the process of integrating the equations, the common important parameter is the time step $\Delta t$.
If a big time step is used, the motion of molecule becomes unstable due to the very big error occurring in the integration.

If a big time step is chosen, the MD
simulation becomes unstable due to the very big error in the integration process, not showin
behaviours existing in the mechanical problem.
If
the time step is very small, the simulation will not be efficient
due to a very long calculation time.

Earlier works ~\cite{Grubmuller1998,Choe2000} have been demonstrated that stable dynamics will be executed only with the use of the smaller time step compared to the period of the highest vibrational frequency. 
Determining the biggest time step for a stable dynamics, will maximize the efficiency of the molecular dynamics simulation.

For these reason and
since every mechanical problem is differnet from each other we decide
to solve the equations for several time steps in order to choose the
indicated such that not take long simulation time but at the same time
we can extract reliable results referring to the lateral force.

\begin{figure}[!h]
  \centering \includegraphics[width =
    1.0\columnwidth]{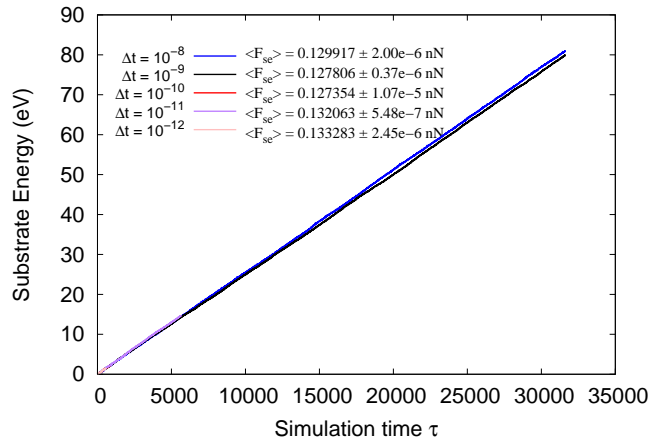}
  \caption{Substrate energy for different different
    time-steps}.\label{Fig:dtcomp1}
\end{figure}

Fig. ~\ref{Fig:dtcomp1} shows the energy of the substrate (kinetic
energy plus elastic potential energy) for differents ${\Delta}t$ :from
0.001 to 10 [ns].  As we are interested in getting the value of
$F_{x}$ from the slope of the curve, it can be observed that the
choose of higher or lower time steps doesnt affect significantly the
final result. The difference can be visualized in graphs such as
position,velocity or $F_{x}$ itself, where for small values of
${\Delta} t$, oscillations of the tip are missing.  In conclusion, for
the numerical simulations we choose ${\Delta} t = 0.1 ns$.

\section{Results}

\begin{figure}[!h]
\centering \includegraphics[width = 1.0\columnwidth]{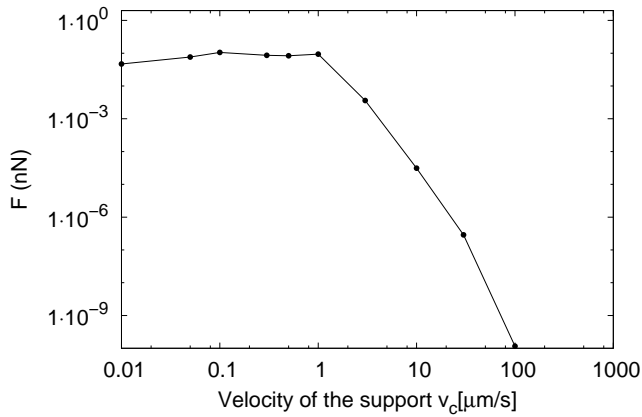}
\caption{Lateral Friction Force in function of the velocity of the
  cantilever. The other parameters were $M = m = 10^{-10}$ $kg$, $K =
k = 10$ $N/m$, $a_{x} = 3$ $\AA$, $U_{0} = 0.085$ $eV$}. \label{Fig:vsvariation}
\end{figure}

Once the method and the time step have been chosen, we calculate $F$ varying
each of the parameters of the model. The variation of $F$ with the velocity of the driven support it is shown in Fig.~\ref{Fig:vsvariation}. For small velocities 
$v_{c} < 1$ $\mu m$, $F$ remains
constant and for higher values, it drops to zero. This is expected
since the model does not have a velocity-dependent damping term.  The
total energy of the substrate decreases as the support velocity
increases because the tip does not interact enough time with the
substrate particles. As a consequence, the amplitude of oscillation around the equilibrium position is not significant.

\begin{figure}[!h]
\centering \includegraphics[width = 1.0\columnwidth]{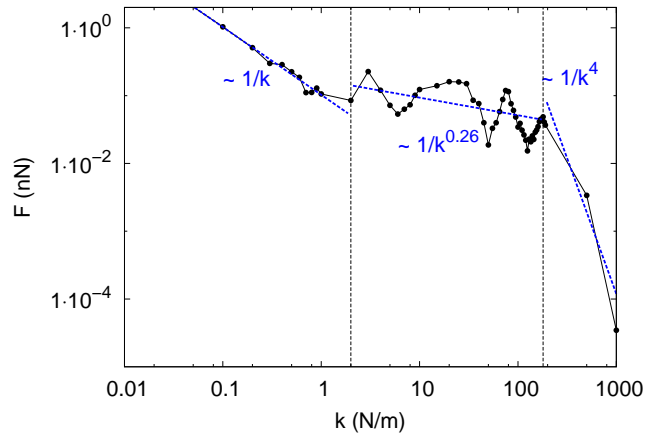}
\caption{Lateral Friction Force with substrate spring stiffness. The other parameters were $M = m = 10^{-10}$ $kg$, $K = 10$ $N/m$, $a_{x} = 3$ $\AA$, $U_{0} = 0.085$ $eV$, $v_{c} = 1$
$\mu m/seg$}. \label{Fig:ksvariation}
\end{figure}

When the variable parameter is the elasticity $k$ of the substrate particles (Fig.~\ref{Fig:ksvariation}), we can identify a behavior of the shape $F = 1/k^{n}$ in three differents regions. 
For the first interval, $k < 1$ $N/m$, the adjustment parameter is $n = 1$. This means that as $k$ increases, $F$ decreases. The softer the material,
the higher the energy lost.
 When 1 $N/m < k < 130$ $N/m$ there is an
unpredictable region with maximums and minimums. Although there is an oscillation in the values of the friction, we assume that the average value remains constant. For that reason, the parameter $n \approx 0.26$.
The third region, shows a more expected result. Due to the hardness of the substrate, the particles do not deviate from their equilibrium position, so the elastic potential is almost null as well as the kinetic energy. In these region, the parameter $n = 4$. 

\begin{figure}[!h]
\centering \includegraphics[width = 0.9\columnwidth]{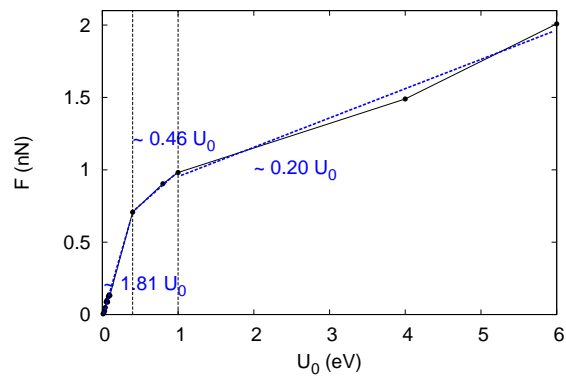}
\caption{Lateral Friction Force in function of the potential high. The other parameters were: $M = m = 10^{-10}$ $kg$, $K =
k = 10$ $N/m$, $a_{x} = 3$ $\AA$, $v_{c} = 1$ $\mu m/seg$}. \label{Fig:u0variation}
\end{figure}

The variation of the friction with the amplitude of the potential $U_{0}$, reproduced in Fig.~\ref{Fig:u0variation}, follows the shape $ F = m U_{0}$, where again $m$ varies for three different regions. In this case, since the data is better visualized and interpreted in linear scale, we did not see the need to represent the results in logarithmic scale.
It can be observed that $F$ increase with the high of potential energy following the shape 
When $U_{0}$ is up to $0.5 eV$, $m  \approx 1.81$. The friction almost doubles its value for each increment in $U_{0}$. For the second region, the growth rate decreases to a quarter of the presented in the first interval, here $m \approx 0.46$. And for the third interval, the growth is already very small, taking into account that it represents the eighth part of the first region growth.
This is because the tip remains in between two particles for more time
(while the support advances), as the amplitude increases.

\begin{figure}[!h]
\centering \includegraphics[width = 0.9\columnwidth]{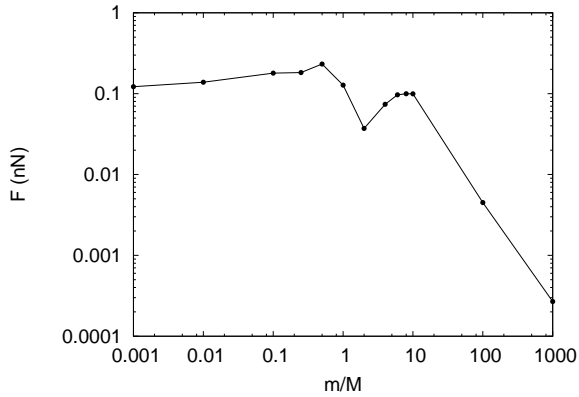}
\caption{Lateral Friction Force in function of the masses ratio. The other parameters were: $K = k = 10$ $N/m$, $a_{x} = 3$ $\AA$, $U_{0} = 0.085$ $eV$, $v_{c} = 1$$\mu m/seg$}. \label{Fig:msvariation2}
\end{figure}

The last Figure correspond to the friction force behavior with
the mass ratio between the particles of the substrate and the tip
$m/M$ (Fig.~\ref{Fig:msvariation2}).
  We can see that when $m < M$ the
frictional force mantains its constant value, then begins to decrease
until reaching a minimum, corresponding to $m = 2 M$.  After this
point increases again and for a small gap of values between $m = 5M$
and $m = 10 M$ again $F$ shows independent on the masses
ratio. Finally, fall to zero for very large substrate masses. When the
nasses are too big, the potential energy of the substrate is very
small since the mass of the tip does not manage to move them.  

\section{Conclusion}

In this paper we investigated a simple model to understand the
fundamental mechanisms on the energy exchange in dry friction without
include any {\em ad hoc} dissipation term as previous works have done.
We compare three different methods to calculate de friction force in order to obtain the most accurate value.
The work was performed solving the Newton's equations numerically, for differents $\Delta t$ with the aim of gain computational time.
We study the variations of $F$ variyng all the parameters involved 
in the description of the model
We could show that the energy lost by the tip is absorbed by the
substrate.
Our gol in the future is to study the same model by placing particles of different masses in the substrate and $T \neq 0$ to obserbe how evolves

\section{Acknowledgments}
This work was supported by the Centro Latinoamericano de Física (CLAF)
and the Conselho Nacional de Desenvolvimento Científico e Tecnológico
(CNPq,Brazil).

\end{document}